# Comprehensive online Atomic Database Management System (DBMS) with Highly Qualified Computing Capabilities

Amani Tahat and Wa'el Salah

Synchrotron-light for Experimental Science and Application in the Middle East (SESAME), Allan 19252, Jordan

**Corresponding author:** "AIMS-Think Energy ", An International Company for Building & Developing Energy Management Information Systems, Amman, 11185 ,Jordan, email: amanitahat@yahoo.com , amanitahat@ AIMS-Think Energy.org

URL: http://www.aims-thinkenergy.org/

### Abstract

*The intensive need of atomic data is expanding continuously in a wide variety of applications (e.g, fusion energy and astrophysics, laser-produced, plasma researches, and plasma processing).This paper will introduce our ongoing research work to build a comprehensive, complete, up-to-date, user friendly and online atomic Database Management System (DBMS) namely called AIMS by using SQLite (http://www.sqlite.org/about.html)(8). Programming language tools and techniques will not be covered here. The system allows the generation of various atomic data based on professional online atomic calculators. The ongoing work is a step forward to bring detailed atomic model accessible to a wide community of laboratory and astrophysical plasma diagnostics. AIMS is a professional worldwide tool for supporting several educational purposes and can be considered as a complementary database of IAEA atomic databases. Moreover, it will be an exceptional strategy of incorporating the output data of several atomic codes to external spectral models.*

## 1. Introduction

This document presents the initial design of our new atomic DBMS. We assume the reader is already familiar with the objectives and architecture of the atomic databases environment. This section presents an introduction of the concept and the need of atomic databases at present. The remainder of this document is organized as follows. Section 2 provides the objectives of this project, besides describing related work and characterizing the difference between the current used atomic databases and AIMS. Section 3 then provides brief description of the online computing capabilities by means of screenshots, functionality as well as performing the validity of the developed calculations. The paper concludes with a summary and discussion of future work in Section 4 and 5.

Nowadays both developing and developed countries are using nuclear based technologies in many industrial applications. Atomic data are required for implementing these technologies. For example, the design and safe operation of modern nuclear installations along with nuclear instrumentation, is only possible on the basis of accurate calculations by using up-to-date nuclear constants so called nuclear and atomic data to be its input. The amount of data needed for such calculations, can be massive. For instance, it is well known that a collection of data for 130 nuclides with more than 10 000 numbers in it must be used for the calculation of the physical behavior of the core of a research reactor as well as its safe operation (http://www-nds.iaea.org/broch.html).Also a set of data of about the same size is used for calculating the radioactive inventory buildup in the reactor and for developing an optimal waste management strategy for the spent fuel of such a reactor. Even more detailed data are mandatory to design an up to date nuclear reactor for the purpose of electricity production and to make decisions on the fuel cycle for today and for the near future. The strict safety regulations must confirm this design and still remain cost effective. The requirements for the quality and accuracy of data for this purpose are very high and need evaluation to be verified and confirmed. On the other hand, radiation therapy of cancer patients is a common application of using nuclear technique in health care. (Electrons, photons, neutrons along with charged particles) are some example of dissimilar types of nuclear radiations that used for this purpose. As a result, the matter of minimizing the damage to the surrounding normal tissue is very important and deepens on the accurateness of the dose delivery at the specified location where it must be better than 5%. In order to determine the dose delivery with such precision, the comprehensive atomic, molecular and nuclear data are required. Someone may ask where I can get such data.

Atomic data can be mainly produced from results of experimental measurements, and software for atomic physics calculations. Laying down on the consideration of each, there are some obstacles. For example, there are several computer programs that can perform atomic calculations. CIV3 [1], SUPER STRUCTURE [2], and the COWAN code [3]. HULLAC [4], ATOM package [5] and the flexible atomic code FAC [6]. Some of these codes are expensive; but most of them are open source and freely distributed for the developments purposes. Although a wide range of codes are exclusive and have never been distributed beyond the producer lab.

On the other hand, consider NIST physics laboratory as a well known atomic database. Yuri Ralchenko (2005) of NIST has kindly provided a description of the new NIST Atomic spectra database which can be found at: http://www.nist.gov/physlab/index.cfm, it contains critically evaluated NIST data for radiative transitions in atomic species, (e.g; Version 3) contains data for the observed transitions of 99 elements and energy levels of 57 elements ). Furthermore, it is an excellent starting place for this kind of material, but the experimental data are not available for all transitions and conditions. In some cases, the theoretical values are preferred to the experimental ones by the NIST evaluators such as calculating the value of oscillator strength gf [7]. However, the process of obtaining experimental data from atomic laboratory still exclusive to certain countries without the other; because they are very expensive and track for political purposes. This underlines the importance of developing the production process of theoretical atomic data and presents how important the subject of this research. Therefore the use of atomic software will be the best way for producing atomic data.

We are doing this project in order to carry out the problems of atomic data recourses as users of atomic data rather than of a producer of atomic data. We carry our responsibility to find the best ways to get accurate, documented, and clear, up-to- date atomic data along with distributing it freely worldwide for scientific research and peaceful purposes. Herby, our group believes that the On-line implementation of an electronic database management system in the company of online computing capability will be an excellent atomic data resource. This contribution will show how the tools of computer aided process engineering can be applied to the optimization of a laboratory measurement.

## 2. The objectives of this project:

Characterizing the difference between the current used atomic databases and AIMS is based on our objectives and they are:

a) The building of AIMS [9], the online Database Management System (DBMS) to facilitate the process of executing queries, data entrance, data manipulation, in addition to sharing data worldwide and effectively contribute the establishment of the complementary atomic database of the currently used atomic database ( e.g., IAEA databases[10] ((http://www-amdis.iaea.org/). Appendix 3 presents more examples of related work.

b) Providing an online competent large set of atomic, and X-ray technology computer programs for generating atomic data. Those programs are proffitional, easy to use and freely distributed. Users can download these programs after his online registration. these sets of programs were created based on:

    a) Some open source atomic codes [6] that are freely distributed and available on the internet for the purpose of evaluation, testing, and development besides some well know physicals formulas. Accordingly, a brief description about each program is presented in section two.

    b) The bulk of AIMS software package based on experimental results that have been conducted through a series of experiments in France laboratories and the Synchrotron-light for Experimental Science and Applications in the Middle East as well. The results were confirmed and published in scientific journals to be addressed as an outcome of continuous work since 2001 in references [11-12].We will describe them in a separated paper.

The currently used atomic database (see appendix 3) are based on an overview of data contained in the database. Using other words, the majority of atomic like IAEA database provides access and search capability for critically evaluated data on some atomic quantities (e.g. energy levels, wavelengths, and transition probabilities that are reasonably up-to-date which is good) and they have gateway to the set of atomic physics codes for providing solutions to any user willing atomic data which can not be easily accessed on the web or which simply do not exist in the database.

For example, the Code Centre Network (CCN) (http://www-amdis.iaea.org/CCN/services/) makes available code capabilities (e.g. online computing and "downloadable" codes ) without creating a database system based on the output generated data . However in AIMS, the DBMS will automatically store the output tabulated data in the database. User can create his own profile and has his own (online or offline) database. Many services are available (e.g. edit, search, save, save as, update, creating a query based on the input parameters, different output files extensions, retrieve data, reset input, online physics consulting …etc). If this user is the system administrator then gusts can only do search inside this database and they are not authorized to generate and/or store data. This new DBMS supported both (online commuting and downloadable codes (offline) .In both cases, user can build his atomic database based on his generated data. This new atomic DBMS has been initially considered as qualified program to register for a software patent in Jordan. More details will be covered in a series of papers after the patent has been issued and is being announced**.**

## 3. Overview of the online computing capabilities:

This project is still under construction. Currently we are ready with two complete online computing programs: the Atomic Structure and Hydrogenic model. Figure 1 and 2 present the main panel of atomic structure window and the Hydrogenic model in the HTML page, respectively. Users can easily switch between them by using the general tab. User manuals are available in two forms HTML and PDF that can be downloaded directly from the website.

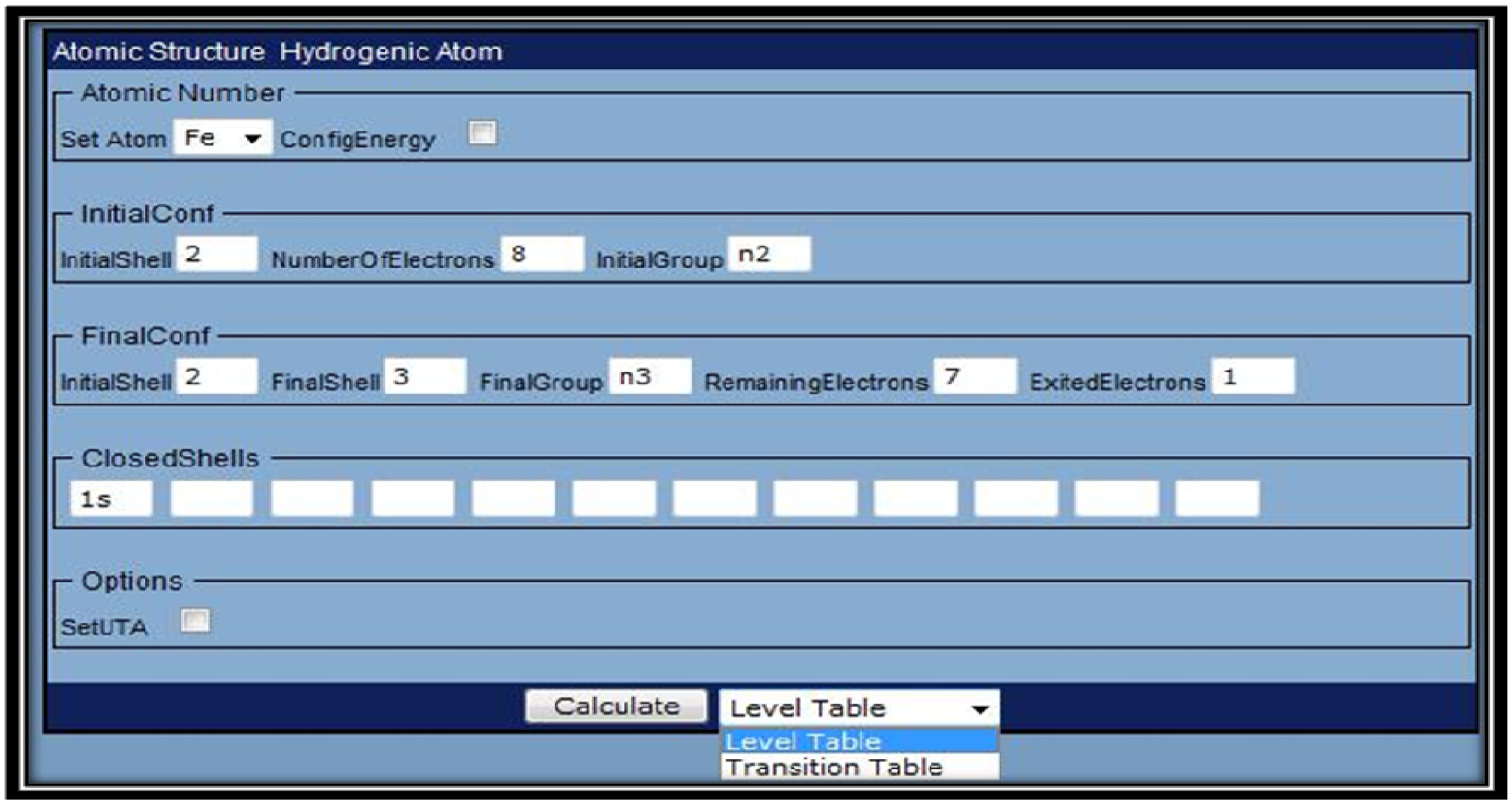

Figure 1: Screenshot of the main panel of atomic structure program which represents the input file.

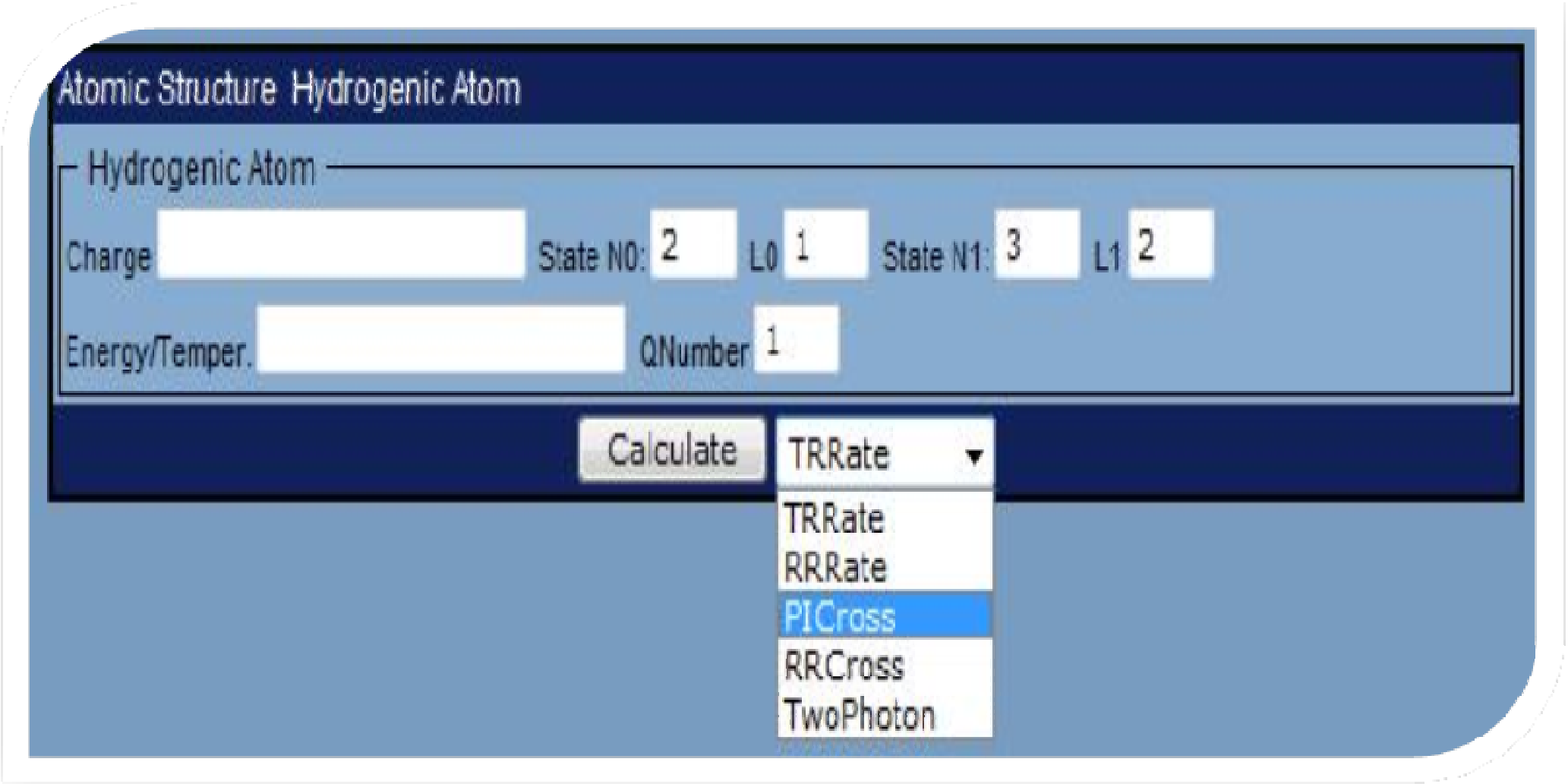

Figure 2: Screenshot of the main panel of Hydrogenic model which represents the input file.

## 3.1 Atomic structure program :

This program computes tabulated atomic structure data, that can be easily interfaced with any other program such as (XSPEC) [13], the output files includes energy levels, oscillator strengths, radiative transition rates, mixing coefficients and reduced multipole matrix elements in addition to a variety of other atomic quantities .The computation process are based on the Configuration Interaction Method (CIM) adopted from the FAC code[6]. It combines the strengths of some available and currently used codes with modifications to numerical methods. User needs to select the option atomic structure from the standard tab, to start working under online dialog mode by filling the input data and choosing the output quantity from the output menu to be calculated on clicking the calculate button in an easy and simple way; user doesn't need to go through the details and complications of the operating systems or the programming languages [14]. Table 2 presents the validity of these calculations. Furthermore the online calculation provides an option (UTA) for improving atomic calculation of wavelengths, oscillator strengths by implementing the so-called unresolved transition array (UTA) which has been identified in many soft X-ray sources as a convenient tool for their interpretation due to low resolution of the soft-X-ray spectra emitted by hot plasmas [15]. In 2001, Behar et al, calculated a complete set of atomic data for such modeling using the Hebrew University Lawrence Livermore Atomic Code (HULLAC) [16], and provided an abbreviated list of transition wavelengths, oscillator strengths, besides some other important atomic quantity. That complete data set has also been included in some commonly used plasma modeling codes. For using this feature user needs to switch to the UTA mode marking the UTA box from option tab (see figure 1). New columns will be added to the output tables to illustrate the UTA effect contains the UTA related transition data. Namely, the transition energy including the UTA shift, the UTA Gaussian width, and the configuration interaction multipler.

| No | Configuration | $J^{\Pi}$ | NIST | AIMS |
|---|---|---|---|---|
| 1 | $1s^{2}_{1/2}2s^{2}_{1/2}$ | $0^{e}$ | 0.0000 | 0.0000 |
| 2 | $1s^{2}_{1/2}\,2s_{1/2}2p_{\,1/2}$ | $0^{o}$ | 21.0528 | 21.2349 |
| 3 | $1s^{2}_{1/2}\,2s_{1/2}2p_{1/2}$ | $1^{o}$ | 21.3431 | 21.5185 |
| 4 | $1s^{2}_{1/2}2s_{1/2}2p_{3/2}$ | $2^{o}$ | 21.9846 | 22.1472 |
| 5 | $1s^{2}_{\ 1/2}\,2s_{\ 1/2}\,2p_{3/2}$ | $1^{o}$ | 40.8750 | 41.8458 |
| 6 | $1s^{2}_{1/2}\,2p^{2}_{1/2}$ | $0^{e}$ | 55.0081 | 55.6194 |
| 7 | $1s^{2}_{1/2}\,2p_{1/2}{}^{2}p_{3/2}$ | $1^{e}$ | 55.3582 | 55.9604 |
| 8 | $1s^{2}_{1/2}{}^{2}p^{2}_{3/2}$ | $2^{e}$ | 55.9125 | 56.5043 |
| 9 | $1s^{2}_{1/2}{}^{2}p_{1/2}{}^{2}p_{3/2}$ | $2^{e}$ | 61.3971 | 62.6257 |
| 10 | $1s^{2}_{1/2}2p^{2}_{3/2}$ | $0^{e}$ | 75.4764 | 77.3522 |
| 11 | $1s_{1/2}2s^{2}_{1/2}2p_{1/2}$ | $0^{o}$ | | 1816.7572 |
| 12 | $1s_{1/2}2s^{2}_{1/2}2p_{1/2}$ | $1^{o}$ | | 1816.8417 |
| 13 | $1s_{1/2}2s^{2}_{1/2}2p_{3/2}$ | $2^{o}$ | | 1817.6862 |
| 14 | $1s_{1/2}{}^{2}s_{1/2}{}^{2}p^{2}_{1/2}$ | $1^{e}$ | | 1821.7062 |
| 15 | $1s_{1/2}2s_{1/2}2p_{1/2}2p_{3/2}$ | $2^{e}$ | | 1822.0689 |
| 16 | $1s_{1/2}2s_{1/2}2p^{2}_{3/2}$ | $3^{e}$ | | 1822.5149 |

Table 2: Target levels of Si xi and their threshold energies (in eV). Portion of table is shown. Complete table will be available electronically in AIMS database. The present results reported here agree with the results compiled in the NIST database [http://www.nist.gov/physlab/data/asd.cfm]. Moreover experimental measurements are not available for all conditions as shown in this table so that the theoretical measurements will be more practice, active and reliable as mentioned in the introduction.

## 3.2 Hydrogenic Ions model:

This online atomic code is planned to be a complete atomic package for calculating various atomic data with an online user friendly interface. This part is mainly provides a hydrogen like ions model by assuming only one electron in the system that has an effective charge state. The theoretical methods used in this model are similar to those Hydrogenic functions of M. F. Gu [17,37] in addition to the functions of Ferland (1996) [18] and Seaton (1959)[19] for calculating the radiative recombination of hydrogenic ions. The hydrogenic model handles mainly the transition probabilities, transition rates, radiative recombination rates, photoionization cross sections, radiative recombination cross sections for the hydrogenic ions beside the calculating of the two-photon decay rate of H-like and He-like transitions 2s1/2 ! 1s1/2 and 1s2sS0 ! 1s2S0. The calculated data will be very useful for astrophysical and plasma's applications due to lack of such data [6, 17, 20].

Hydrogenic Ions model user guide [21] presents more explanation (e.g., describing input files, output files, clarifications of the parameters that must be defined by user in order to use this model online by providing some important screenshots of the main panels of this model , along with providing the validity of it.

AIMS provides a gate for obtaining more accurate results to get the best atomic data by implementing additional formulas for calculating recombination rate coefficients and other atomic quantities to allow comparing results. Data accuracy is recommended in extensively astrophysical applications. These formulas are:

1. The (D. A. Verner & G. J. Ferland) 1996 formulas for H-like ions besides He-like, Li-like and Na-like ions over a broad range of temperature [22]. Theory and fitting parameters as well as limitations are available to public in the web site http://www.pa.uky.edu/~verner/rec.html.
2. Seaton (1959): For the hydrogenic species, Arnaud & Rothenflug (1985) recommended this formula but it is not valid at high tempretures $T > 10^6 Z^2$. See reference ( D. A. Verner and G. J. Ferland,1995).
3. Power-law fits to the rates of radiative recombination. S. M. V. Aldrovandi & D. Pequignot[23 ], J. M. Shull & M. Van Steenberg [24 ], M. Arnaud & R. Rothenflug [25 ].We create our c++ function to calculate the radiative recombination rates by using **the following fitting formula** $\alpha_r(T) = A(T/10^4)^{-B}$ **Available with its fitting parameters A and B , on ftp://gradj.pa.uky.edu//dima//rec//pl.txt .**Table 3 presents a comparison between the current results and those from literate.AIMS shows a good agreements with them.

| **Ion** | **AIMS** | **Gould(1978)** | **[a]Old fit** |
|---|---|---|---|
| Si IV | $7.015029 \times 10^{-12}$ | $7.32 \times 10^{-12}$ | $5.5 \times 10^{-12}$ |
| C IV | $8.202417 \times 10^{-12}$ | $8.45 \times 10^{-12}$ | $9.16 \times 10^{-12}$ |
| Mg II | $1.215735 \times 10^{-12}$ | $1.35 \times 10^{-12}$ | $8.8 \times 10^{-13}$ |

Table 3: Radiative recombination coefficients (in $cm^3 s^{-1}$), at $T= 10^4$K, [a] Arnaud & Rothenflug ( 1985) for C IV , Aldrovandi & Pequignot ( 1973) for Mg II and Si IV.

| **Temperature (k)** | **RRRate coefficient $Cm^3\ s^{-1}$ Quantum number = 1 (Gu,2008)** | **RRRate coefficient $Cm^3\ s^{-1}$ Quantum number = 1 D. A. Verner** | **RRRate coefficient $Cm^3\ s^{-1}$ Quantum number = 1 Seaton** |
|---|---|---|---|
| 3 | $2.4859 \times 10^{-8}$ | $2.4027 \times 10^{-8}$ | $2.9974 \times 10^{-8}$ |
| 3.5 | $2.3013 \times 10^{-8}$ | $2.2215 \times 10^{-8}$ | $2.7505 \times 10^{-8}$ |
| 4 | $2.1525 \times 10^{-8}$ | $2.0754 \times 10^{-8}$ | $2.5530 \times 10^{-8}$ |
| 4.5 | $2.0293 \times 10^{-8}$ | $1.9544 \times 10^{-8}$ | $2.3904 \times 10^{-8}$ |
| 5 | $1.9250 \times 10^{-8}$ | $1.8521 \times 10^{-8}$ | $2.2537 \times 10^{-8}$ |
| 5.5 | $1.8352 \times 10^{-8}$ | $1.7641 \times 10^{-8}$ | $2.1367 \times 10^{-8}$ |
| 6 | $1.7570 \times 10^{-8}$ | $1.6873 \times 10^{-8}$ | $2.0352 \times 10^{-8}$ |
| 6.5 | $1.6879 \times 10^{-8}$ | $1.6196 \times 10^{-8}$ | $1.9460 \times 10^{-8}$ |
| 7 | $1.6264 \times 10^{-8}$ | $1.5593 \times 10^{-8}$ | $1.8668 \times 10^{-8}$ |
| 7.5 | $1.5711 \times 10^{-8}$ | $1.5051 \times 10^{-8}$ | $1.7960 \times 10^{-8}$ |
| 8 | $1.5211 \times 10^{-8}$ | $1.4561 \times 10^{-8}$ | $1.7322 \times 10^{-8}$ |
| 8.5 | $1.4755 \times 10^{-8}$ | $1.4114 \times 10^{-8}$ | $1.6743 \times 10^{-8}$ |
| 9 | $1.4339 \times 10^{-8}$ | $1.3706 \times 10^{-8}$ | $1.6214 \times 10^{-8}$ |
| 9.5 | $1.3955 \times 10^{-8}$ | $1.3330 \times 10^{-8}$ | $1.5730 \times 10^{-8}$ |
| $10^1$ | $1.3601 \times 10^{-8}$ | $1.2983 \times 10^{-8}$ | $1.5283 \times 10^{-8}$ |
| $10^2$ | $4.2406 \times 10^{-9}$ | $3.8814 \times 10^{-9}$ | $4.1479 \times 10^{-9}$ |
| $10^3$ | $1.1886 \times 10^{-9}$ | $1.0847 \times 10^{-9}$ | $1.0957 \times 10^{-9}$ |
| $10^4$ | $2.1225 \times 10^{-10}$ | $2.7956 \times 10^{-10}$ | $2.7870 \times 10^{-10}$ |
| $10^5$ | $1.9210 \times 10^{-11}$ | $6.7295 \times 10^{-11}$ | $6.7021 \times 10^{-11}$ |
| $10^6$ | $0.9453 \times 10^{-11}$ | $1.4993 \times 10^{-11}$ | $1.4742 \times 10^{-11}$ |
| $10^7$ | $1.1827 \times 10^{-12}$ | $2.8404 \times 10^{-12}$ | $2.7752 \times 10^{-12}$ |
| $10^8$ | $-0.0780 \times 10^{-13}$ | $3.7748 \times 10^{-13}$ | $3.8534 \times 10^{-13}$ |
| $10^9$ | $-0.1580 \times 10^{-14}$ | $2.9238 \times 10^{-14}$ | $3.7306 \times 10^{-14}$ |

Table 4: present a comparison between the calculated values of the recombination rate (RRRate) coefficient of Ci XVII, by using three formulas Gu, Verner, Seaton.

- ***Results annalysis :***

It is extremely important; among other things to know with certainty the radiative recombination coefficient rate. Detecting the accuracy is recommended in all energy, plasma, atomic researches as well as atmosphere studies .For example; researchers may need to calculate the rate of radiative recombination for some atoms along with ions of the upper atmosphere. In that case, our program provides three options to investigate the suitable value when computing accurate rate coefficients for the radiative recombination. Simple expressions are also presented for their quick estimation far away from losing time when trying to compare the obtained results with other programs results. In general, as shown in the figure (3) below all methods behave the same behavior, as its producing results show the same behavior depending on the temperature rising or declining. More precisely, the selection mechanism of the value at a certain temperature; extremely affects the results in a very large scale in terms of atmosphere studies language**.**

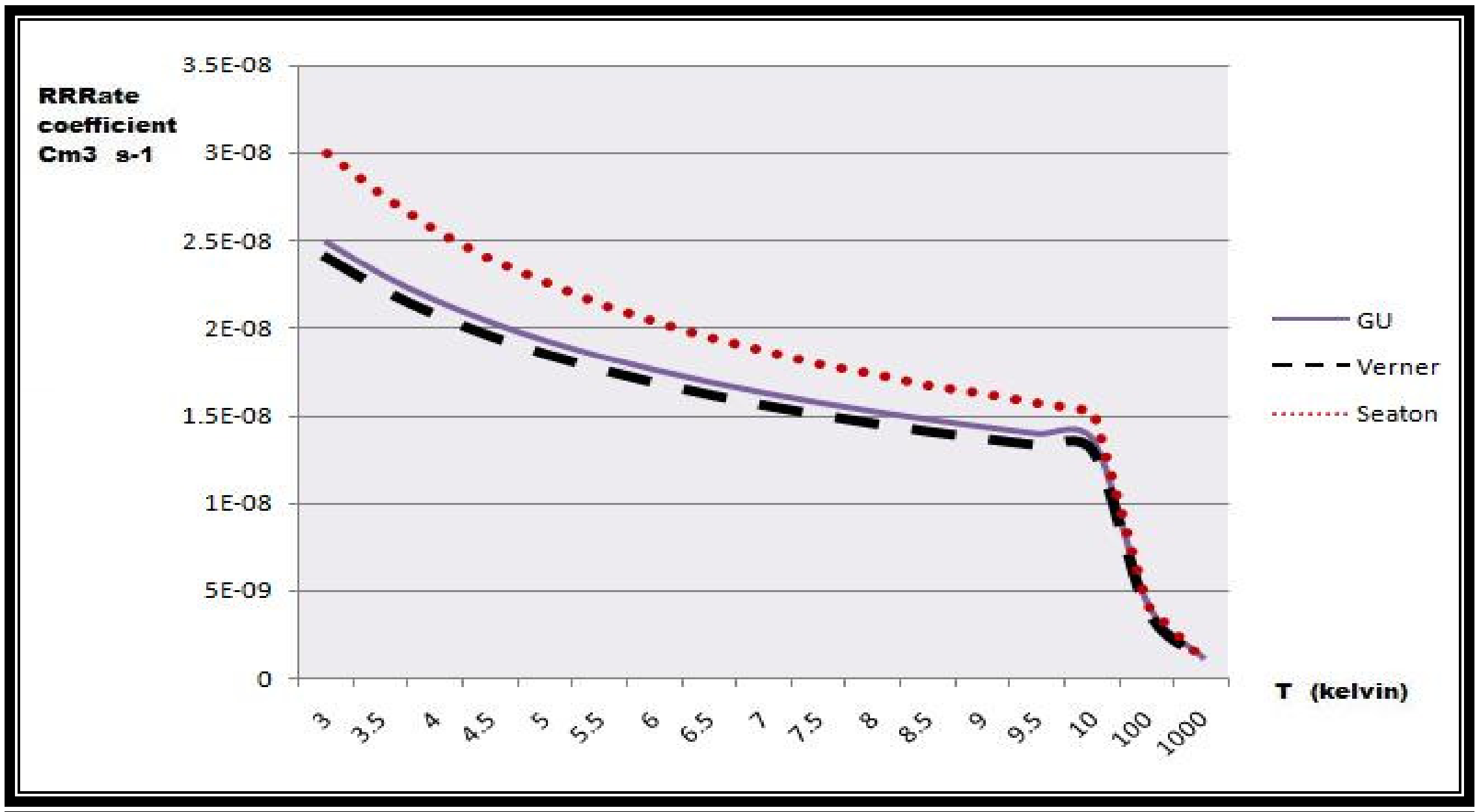

Figure 3 : RRRate coefficient versus tempreture ,by using Gu,Verner and Seaton

### 3.3 Spectroscopy:

This part is a very important part of the created database. Because the convenient access and the effective use of atomic physics data have long been goals of X-ray spectroscopy researches. Additionally, the very considerable advances in computing technology now allow the use of extensive radiative models. Unfortunately, the goal of convenient access and effective use of atomic data remains hard to pin down. Although the propagation of computing capabilities has made it progressively easier to carry out large-scale atomic rate calculations, these advances have been offset by difficulties in managing and evaluating the data. For the purposes of understanding, it is important that the data which goes into the model atoms could be readily examined, compared and manipulated. We are undertaking to develop a set of computational capabilities which will facilitate the access, manipulation, and understanding of atomic data in calculations of x-ray spectral modeling to be worked online from AIMS website. In this present limited description we will emphasize the objectives for this part of the whole project, the design philosophy, and aspects of the atomic database. A complete description of this work is avai1able in [26].

Originally, spectroscopy is the study of the interaction between radiation and matter as a function of wavelength ($\lambda$). Common types of it are Absorption and Fluorescence. X-ray spectroscopy can be considered as a gathering name for several spectroscopic tools that are usually used for determining the electronic structure of materials by using x-ray excitation. Interaction between accelerated charged particles and matter leads to the emission of characteristic x-rays. X-Ray Fluorescence, which is also known as Secondary interelement. Fluorescence occurs where the secondary X-rays emitted by a heavier element are sufficiently energetic to stimulate additional secondary emission from a lighter element. This phenomenon can also be modeled, and corrections can be made providing that the full matrix composition can be deduced. It is also possible to create a characteristic secondary X-ray emission using other incident radiation to excite the sample: 1) electron beam: electron microprobe or Castaing microprobe, 2) ion beam: particle induced X-ray emission (PIXE), where particle-induced X-ray emission (PIXE) is a well known analytical tools for the determination of elemental concentrations of various types of samples of biological, geological, archaeological or environmental nature. On the other hand, secondary X-ray Fluorescence (XRF) and Synchrotron Radiation are very important related topics when it comes to utilize the x-rays that produced by using Electron Probe Microanalysis (EPMA) for many research applications. In particular:

- XRF (X-Ray Fluorescence), where x-rays from (a sealed tube) are used to produce x-rays by secondary fluorescence in samples of interest. Usually it will be a macro-technique.
- Synchrotron Radiation, where electrons are accelerated in ~10s-100s meters diameter rings, and then made to produce highly focused beams of extremely intense x-rays or light, which will then fed into many different types of experiments.
- The benefits of secondary x-ray fluorescence include very low detection limits (10s of ppm easy in 10 seconds, no backgrounds)

In this part of AIMS we designed, wrote and examined several PIXE spectra in a utility program namely called WPASS-2[26, 27, and 28] based on the existing PIXE analysis software package PIXAN [29] with new features that consider the secondary interelement fluorescence in order to improve the basic capabilities and accuracy of PIXAN. Development of such programs handles a few computer codes, is quite meaningful.

Series of papers have been published by the authors to evaluate the academic values of this program as well as describing the PIXE method that used for the analysis of PIXE spectra [28, 27]. Moreover, in order to handle the matter of secondary interelement fluorescence of PIXE..etc, we create a model namely called integrals [30] for presenting an exact but computationally treatment of Quantitative X-ray fluorescence analysis using the fundamental parameter method of sparks (1975) [31], where physical parameters such as photoelectric cross-sections, total mass attenuation cross-sections, fluorescence yields, X-ray branching ratio, detector efficiency (calculated theoretically) are given as inputs. The program then theoretically generates fluorescent X-rays and mainly calculate Secondary enhancement (intensity).

Fluorescent intensity as well as that excited by the incident radiation may be excited by the fluorescence from other elements in the sample .For example , in Ni-fe alloys the Ni K radiation is of sufficient energy to excite Fe K radiation giving an additional intensity of Fe K radiation ,$\Delta I_{FeK}$ , over that calculated equation 17 ,see figure 4. Ignoring these enhancement effects would limit the application of x-ray florescence to bulk samples comprising a few elements widely separated in atomic number, to samples composed only of trace elements and to very thin samples such as air filters where enhancement effects are negligible because of specimen thinness.

from element i caused by the element j is defined as $(\Delta I/I)_{i,j}$ and given by

$$(\Delta I/I)_{i,j} = \frac{C_j[(\mu_k/\rho)_o \omega_k f_k]_j [(\mu_k/\rho)_j]_i}{2[(\mu_k/\rho)_o]_i} [\sin\psi/(\mu_{s,o}/\rho_s)] \ln(\frac{\mu_{s,o}/\sin\psi}{\mu_{s,j}} + 1)$$

$$+ \sin\phi/(\mu_{s,i}/\rho_s) \ln(\frac{\mu_{s,i}/\sin\phi}{\mu_{s,j}} + 1)] , \quad (17)$$

Figure 4 : Equation 17 adapted from sparks(1975)[34].

Our model employs the fundamental parameter method for quantitative analysis by iteration via two easy to use interfaces (customize and general) by choosing the specified interface depended on the available parameters (see appendix 3 and the manual [26]). The model is described for quantitative energy dispersive X-ray fluorescence analysis. We compare the present results with (sparks, 1975) results to show the validity of our model, since this work provides a computational solution for the same method of sparks.

Given in appendix 1 are necessary parameters and results of a calculation using equation 17 for the secondary enhancement of $FeK_\alpha$ radiation resulting from $Nik_\alpha$ and $NiK_\beta$ radiation in $Ni - Fe$ alloys excited by $M_o k_\alpha$ radiation .The comparison of our results and those of sparks is presented in table 5,and it is based on the information from both (appendix1 and appendix 2) .

| **Secondary enhancement by** $Nik_{\alpha}$ | | **Secondary enhancement by** $NiK_{\beta}$ | | **Total** $Nik_{\alpha} + K_{\beta}$ | |
|---|---|---|---|---|---|
| $(\Delta I / I)_{FeK\alpha, NiK\alpha}$ | | $(\Delta I / I)_{FeK_{\alpha}, NiK_{\beta}}$ | | $(\Delta I / I)_{FeK_{\alpha}, NiK}$ | |
| SPARKS | AIMS | SPARKS | AIMS | SPARKS | AIMS |
| 0.019 | 0.049 | 0.002 | 0.032 | 0.021 | 0.051 |
| 0.039 | 0.069 | 0.009 | 0.039 | 0.044 | 0.074 |
| 0.085 | 0.115 | 0.011 | 0.041 | 0.096 | 0.126 |
| 0.140 | 0.170 | 0.018 | 0.048 | 0.158 | 0.188 |
| 0.206 | 0.236 | 0.027 | 0.057 | 0.232 | 0.262 |
| 0.288 | 0.318 | 0.037 | 0.067 | 0.325 | 0.355 |
| 0.393 | 0.423 | 0.050 | 0.080 | 0.443 | 0.473 |
| 0.531 | 0.561 | 0.068 | 0.098 | 0.599 | 0.629 |
| 0.723 | 0.753 | 0.091 | 0.121 | 0.815 | 0.845 |
| 1.008 | 1.038 | 0.126 | 0.156 | 1.134 | 1.164 |

Table (5): The comparison of our finding and those of sparks, 1975 (fundamental method).

## ❖ Comparison analysis:

The shortcomings of reliability coefficients in most statistical studies suggest that a more useful measure of test reliability is to be found in the standard error of measurement SEM (See, for example, Norcini, 2000) [32] test performance .Furthermore, The standard error of measurement (SEM) is an estimate of error to be used in interpreting an individual's test score where the test score is an estimate of a person's "true" by using a reliability coefficient and the test's standard deviation. The (SEM) is helpful for quantifying the margin of error that occurs on every test. Hereby, the current comparison analysis attempt to calculate SEM based on "the Standard Error of Measurement module "that has been produced by (Harvill (1991) [33] including the theory of mental tests of Gulliksen (1950) [34] as discussed in Harvill (1991). However, the data in Appendix 2 are real performance data and has been generated by using: 1) generated data of Sparks's fundamental parameters; 2) generated data of AIMS.

Based on the foregoing; we can read the following from appendix 2.

1. The table has two columns give the total enhancement of $FeK_{\alpha}$ intensity caused by both the $Nik_{\alpha}$ and $NiK_{\beta}$ radiations from both Sparks and AIMS.
2. Two sample sizes =10 (each sample consists of 10scores).
3. Reliability indicates the degree to test scores to which they are free of errors of measurement. Sparks score and AIMS score variances are equal, then the test reliability equals the highest possible value (r =1), this leads to (SEM) value equals Zero (SEM=0). Our finding means that test is perfectly reliable, scores are consistent, dependable, repeatable, that is, no errors of measurements with perfectly reliable test ,a set of errors all equal to zero has no variability .
4. Standard deviation values of both (sparks and AIMS) tests are equaled and have the value of $\delta = 0.34725$ , but the mean of sparks scores is greater than the mean of AIMS scores they are (0.3867 and0.4167 )respectively .Furthermore , The mean describes what is being measured, while the standard deviation represents noise and other interference. In these cases, the standard deviation is not important in itself, but only in comparison to the mean. In some situations, two other terms need to be mentioned: 1) signal-to-noise ratio (SNR), which is equal to the mean divided by the standard deviation, 2) the coefficient of variation (CV). This is defined as the standard deviation divided by the mean, multiplied by 100 percent. Better data means a higher value for the SNR and a lower value for the CV. Implementing this issue to our calculation we can read the following:

| **Index** | **Sparks** | **AIMS** |
|---|---|---|
| **Standard deviation** | 0.34725 | 0.34725 |
| **Mean** | 0.3867 | 0.4167 |
| **SNR** | 1.1 | 1.2 |
| **CV** | 0.89% | 0.83% |

Generally , SNR'S and CV'S values are very close to each other this indicate that both sparks and AIMS tests have a better and performance data where each of them has a higher value for the SNR and a lower value for the CV. After all, using AIMS will be a good strategy for reexamining SPARK's measured values easily with a very short time. The following two examples can explain the idea:

1) For instance, comparing appendix 2 with appendix 1 ,user need to use the customer interface of this model to compose his input file by fitting the entire fundamental parameters and setting them to those of sparks from the main panel of model,see appendix 3.In the main time user can avoid details and use the general interface that has 7 parameters each of them consist of a set of some other parameters like the angles($\psi_1,\psi_2$), $C_{Ni}$, $C_{Fe}$, $\mu$ , $\rho$ …etc,

2) Calculating the secondary enhancement correction as a function of the angles ($\psi_1,\psi_2$),shows the dependence to become significant when the enhancement is large see appendix 1.In AIMS any arbitrary input can proof this property and others by furthermore ,in such cases AIMS has two limitations

When $h \to 0 \Rightarrow S \to 0$; where S is the intensity

When $h \to \infty \Rightarrow S \to A6\,K$, These two conditions has been applied perfectly with what has been explained in the theory of Spark. Finally Sparks measurements has been evaluated and tested by based on (Shiraiwa and Fujino, 1966) [35] by using the same factors with equation 17 and presents a good agreement with a results published by (Muller). Consequently, this can improve the validity of using AIMS as well as raise its accuracy.

## 4. Future work:

We are working on adding a program for x-ray microprobe imaging and allow it to be worked online, besides adding the following atomic programs:

1. colliosonal ionaization model
2. Collision Strength Measurement for Electron-Impact Excitation model
3. Radiative recombination and photoionization model **.**

## 5. Summary and Conclusion

We have successfully proposed and designed an online database management system namely called AIMS, with current and anticipated communication needs in mind. By using fully scalable architecture optimized for speed and accuracy that manages and support efficient access to very large amount of data and support concurrent access to online computing models. Thus provide full service web-based solutions. Following are some of our most requested online services:

- Dynamic Website Design
- Custom programming contains several computer models for producing atomic data based on the most reliable atomic theories besides some experimental models .user can use them online or download them freely on his personal computer.
- Database creation and integration(Expandability ),Data Collection(Data Entry, Batch Controls, file enhancements, file verification)**,**Output: Queries and Profiles
- Internet atomic physics online consulting

As atomic research grows, demands on atomic database will be greater. Physicists need to make sure that the data going into the on-line database are clean and accurate. They want to make sure the product has the functionality they are looking for now and in the future. They want to make sure that our data are safe. For this reason, we developed a series of computational models which were convincing to physicists and could quantify the performance advantages of using database and object management technologies. These models are :Atomic structure program ,Hydrogenic model, spectroscopy model including wpass program and Several models based on our experimental result .Also we have described and verified the validity of our system by comparing the analysis of atomic physics data using a AIMS with NIST , Sparks ,Seaton and D. A. Verner as well as ming fing Gu data. Our results show a good agreements with literature data ,for instance ,when finding the standard error of measurement (SEM) for the spectroscopy model ,Sparks score and AIMS score variances were equaled , and test reliability equaled the highest possible value (r =1), with (SEM) equals Zero (SEM=0).thus means that our test result are perfectly reliable, scores are consistent, dependable, repeatable, that is, no errors of measurements once we have got the perfectly reliable test .

With the proofing -of-our concept system, we demonstrated that analysis of atomic physics data could be completed about an order of magnitude faster than with existing production codes where most of them are expensive and exclusive to their producers .Furthermore, AIMS-DBMS can be considered as a complementary database of the currently used atomic database such as the IAEA databases as well as working as a professional worldwide tool for supporting several educational purposes. Finally AIMS package is freely distributed online from its official website and still under construction. Many publications about the AIMS theoretical background and programming languages have been published and mentioned as references during the bulk of the research.
This project was an initiative of us purely for scientific purposes and we continue the development and presentation of information for free, but given the importance of the subject matter, and the world urgent need for such a project, we make a public call for all interested to expand this work by supporting us with scientific participations.

## Appendix 1

| | Sparks | | AEMIS | | | | | |
|---|---|---|---|---|---|---|---|---|
| | $(\Delta I/I)_{FeK_\alpha, NiK}$ | $(\text{Variance})^2$ | $(\Delta I/I)_{FeK_\alpha, NiK}$ | $(\text{Variance})^2$ | Test reliability Of Measurements | $S_E^2$ | Standard Error Measurements(SEM) | Test reliability Among Erorr scores |
| | Total enhancement of $FeK_\alpha$ Caused by both, $Nik_\alpha$ & $NiK_\beta$ radiations | $S_T^2$ | Total enhancement of $FeK_\alpha$ Caused by both, $Nik_\alpha$ & $NiK_\beta$ radiations | $S_X^2$ | $r_{xx'} = \frac{S_T^2}{S_X^2}$ | $S_E^2 = S_X^2(1-r_{xx'})$ | $S_E = \sqrt{S_X^2(1-r_{xx'})}$ | $r_{xx'} = 1 - \frac{S_E^2}{S_X^2}$ |
| | 0.021 | 0.13373649 | 0.051 | 0.13373649 | 1 | 0 | 0 | 1 |
| | 0.044 | 0.11744329 | 0.074 | 0.11744329 | 1 | 0 | 0 | 1 |
| | 0.096 | 0.08450649 | 0.126 | 0.08450649 | 1 | 0 | 0 | 1 |
| | 0.158 | 0.05230369 | 0.188 | 0.05230369 | 1 | 0 | 0 | 1 |
| | 0.232 | 0.02393209 | 0.262 | 0.02393209 | 1 | 0 | 0 | 1 |
| | 0.325 | 0.00380689 | 0.355 | 0.00380689 | 1 | 0 | 0 | 1 |
| | 0.443 | 0.00316969 | 0.473 | 0.00316969 | 1 | 0 | 0 | 1 |
| | 0.599 | 0.04507129 | 0.629 | 0.04507129 | 1 | 0 | 0 | 1 |
| | 0.815 | 0.18344089 | 0.845 | 0.18344089 | 1 | 0 | 0 | 1 |
| | 1.134 | 0.55845729 | 1.164 | 0.55845729 | 1 | 0 | 0 | 1 |
| Arithmetic Mean $\bar{X}$ | 0.3867 | 0.12058681 | 0.4167 | 0.12058681 | | | | |
| Standard Deviation $\delta$ | 0.34725 | | 0.34725 | | | | | |

# Appendix2

Table 2. Calculation of Secondary Enhancement Correction, $\Delta I/I$, to the Fe $K_\alpha$ Intensity Resulting from Ni $K_\alpha$ and Ni $K_\beta$ Radiation in Ni-Fe Alloys Excited by Mo $K_\alpha$ Radiation

| | | Secondary Enhancement by Ni $K_\alpha$ | | | | | | Secondary Enhancement by Ni $K_\beta$ | | | Total Ni $K_\alpha$ + $K_\beta$ |
|---|---|---|---|---|---|---|---|---|---|---|---|
| | | | | | $(\Delta I/I)_{Fe\ K_\alpha, Ni\ K_\alpha}$ | | | | | | |
| $C_{Ni}$ | $C_{Fe}$ | $C_{Ni}A/2$ $cm^2/g$ | $\mu_{s,c}/\rho_s$ $cm^2/g$ | $\mu_{s,j}/\rho_s$ $cm^2/g$ | $\psi=\phi=45°$ | $\psi=\phi=20°$ | $\psi=30°$ $\phi=60°$ | $C_{Ni}B/2$ $cm^2/g$ | $\mu_{s,j}/\rho_s$ $cm^2/g$ | $(\Delta I/I)_{Fe\ K_\alpha, Ni\ K_\beta}$ $\psi=\phi=45°$ | $(\Delta I/I)_{Fe\ K_\alpha, Ni\ K}$ $\psi=\phi=45°$ |
| .05 | .95 | 3.63 | 37.0 | 351.2 | 0.019 | | | 0.372 | 271.5 | 0.002 | 0.021 |
| .10 | .90 | 7.26 | 37.4 | 335.8 | 0.039 | 0.036 | 0.039 | 0.744 | 259.6 | 0.005 | 0.044 |
| .20 | .80 | 14.52 | 38.3 | 305.1 | 0.085 | | | 1.488 | 235.8 | 0.011 | 0.096 |
| .30 | .70 | 21.78 | 39.2 | 274.5 | 0.140 | | | 2.232 | 212.0 | 0.018 | 0.158 |
| .40 | .60 | 29.04 | 40.1 | 243.8 | 0.206 | | | 2.976 | 188.1 | 0.027 | 0.232 |
| .50 | .50 | 36.30 | 41.1 | 213.2 | 0.288 | 0.252 | 0.286 | 3.720 | 164.3 | 0.037 | 0.325 |
| .60 | .40 | 43.55 | 42.0 | 182.5 | 0.393 | | | 4.464 | 140.5 | 0.050 | 0.443 |
| .70 | .30 | 50.81 | 42.9 | 151.8 | 0.531 | | | 5.208 | 116.7 | 0.068 | 0.599 |
| .80 | .20 | 58.07 | 43.8 | 121.1 | 0.723 | | | 5.952 | 92.87 | 0.091 | 0.815 |
| .90 | .10 | 65.33 | 44.7 | 90.5 | 1.008 | 0.796 | 0.993 | 6.697 | 69.06 | 0.126 | 1.134 |

where $A = [(\mu_K/\rho)^a_{Mo\ K_\alpha}\ \omega^d_K f^d_{K_\alpha}]_{Ni}\ [(\mu_K/\rho)_{Ni\ K_\alpha}]^a_{Fe}/[(\mu_K/\rho)_{Mo\ K_\alpha}]^a_{Fe} = 40.69\ cm^2/g \times 0.414 \times 0.881$

$\times\ 319.71\ cm^2/g + 32.68\ cm^2/g = 145.19\ cm^2/g,$

$\mu_{s,c}/\rho_s = C_{Ni}\ (\mu/\rho)^a_{Ni,Mo\ K_\alpha} + C_{Fe}\ (\mu/\rho)^a_{Fe,Mo\ K_\alpha} = C_{Ni} \times 45.6\ cm^2/g + C_{Fe} \times 36.5\ cm^2/g,$

$\mu_{s,j_{K_\alpha}}/\rho_s = C_{Ni}\ (\mu/\rho)^c_{Ni,Ni\ K_\alpha} + C_{Fe}\ (\mu/\rho)^c_{Fe,Ni\ K_\alpha} = C_{Ni} \times 59.8\ cm^2/g + C_{Fe} \times 366.5\ cm^2/g,$

$B = [(\mu_K/\rho)^a_{Mo\ K_\alpha}\ \omega^d_K f^d_{K_\beta}]_{Ni}\ [(\mu_K/\rho)_{Ni\ K_\beta}]^a_{Fe}/[(\mu_K/\rho)_{Mo\ K_\alpha}]^a_{Fe} = 40.69\ cm^2/g \times 0.414 \times 0.119$

$\times\ 242.6\ cm^2/g + 32.68\ cm^2/g = 14.88\ cm^2/g,$

and $\mu_{s,j_{K_\beta}}/\rho_s = C_{Ni}\ (\mu/\rho)^c_{Ni,Ni\ K_\beta} + C_{Fe}\ (\mu/\rho)^c_{Fe,Ni\ K_\beta} = C_{Ni} \times 45.24\ cm^2/g + C_{Fe} \times 283.4\ cm^2/g.$

a. Cromer and Liberman (5), c. McMasters et al. (8,9), d. Bambynek et al. (10).

## Appendix 3

| **Name** | **Website** |
|---|---|
| The NIST Atomic Spectra Database | **http://www.nist.gov/pml/data/asd.cfm** |
| **NIFS Database**<br>Atomic & Molecular Numerical Databases | **http://dbshino.nifs.ac.jp/** |
| **LAMDA**<br>Leiden atomic and molecular Database | **http://www.strw.leidenuniv.nl/~moldata/** |
| **CHIANTI**<br>An Atomic Database for Spectroscopic Diagnostics of Astrophysical Plasmas | **http://www.chianti.rl.ac.uk/** |
| **ALADDIN Database**<br>Electron Collisions<br>Heavy Particle Collisions | **http://www-amdis.iaea.org/ALADDIN/** |
| **AMBDAS**<br>Atomic and Molecular Bibliographic Data System | **http://www-amdis.iaea.org/AMBDAS/** |
| ***Atomic Database***<br>Atomic & Molecular Database | **http://www.camdb.ac.cn/e/** |